\documentclass[aps,12pt,nofootinbib]{revtex4}
\usepackage{epsf,amsfonts,amsmath}
\usepackage{graphicx,float,subfigure,epsfig,color,rotating,latexsym}
\usepackage{comment}
\usepackage{epsfig}

\def\beq{\begin{equation}}
\def\eeq{\end{equation}}

\def\bea{\arraycolsep .1em \begin{eqnarray}}
\def\eea{\end{eqnarray}}
\def\Tr{{\rm Tr}}
\newcommand{\step}{\vspace{.5em}}

\def\s0#1#2{\mbox{\small{$ \frac{#1}{#2} $}}}
\def\0#1#2{\frac{#1}{#2}}

\def\grgl{\:\hbox to -0.2pt{\lower2.5pt\hbox{$\sim$}\hss}{\raise3pt\hbox{$>$}}\:}
\def\klgl{\:\hbox to -0.2pt{\lower2.5pt\hbox{$\sim$}\hss}{\raise3pt\hbox{$<$}}\:}

\newcommand \be {\begin{equation}}
\newcommand \ee {\end{equation}}
\newcommand \bed {\begin{displaymath}}
\newcommand \eed {\end{displaymath}}

\newcommand{\bit}{\begin{itemize}}
\newcommand{\eit}{\end{itemize}}


\begin{document}
\graphicspath{{FIGURE/}}

\title{A compared analysis of the susceptibility in the $O(N)$ theory}

\author{Vincenzo Branchina, Emanuele Messina}

\affiliation{
\mbox{Dipartimento di Fisica, Universit\`a di Catania, 64 via S. Sofia, I-95123, Catania, Italy;}\\
{INFN, Sezione di Catania, 64 via S. Sofia, I-95123, Catania, Italy }
}%

\author{Dario Zappal\`a}

\affiliation{
\mbox{INFN, Sezione di Catania, 64 via S. Sofia, I-95123, Catania, Italy.}
}%

\begin{abstract}
${}$\\[-1ex]
\centerline{\bf Abstract\hskip 90 pt}
The longitudinal  susceptibility $\chi_L$ of the $O(N)$ theory in the broken phase  is analyzed by means of three different approaches,
namely the leading contribution of the $1/N$ expansion, the Functional Renormalization Group flow  in the Local 
Potential approximation and the improved effective potential via the Callan-Symanzik equations, properly extended to 
$d=4$ dimensions through the  expansion in powers of $\epsilon=4-d$.  The findings of the three approaches are compared 
and their agreement in the large $N$ limit is shown.  The numerical analysis of the Functional Renormalization Group flow equations
at small $N$ supports the vanishing of  $\chi_L^{-1}$ in $d=3$ and $d=3.5$ but is not conclusive in $d=4$,
where we have to resort to  the  Callan-Smanzik approach.  At finite $N$ as well as in the 
limit  $N\to\infty$, we find that $\chi^{-1}_L$ vanishes with $J$
as $J^{\epsilon/2}$ for $\epsilon>0$ and as $(\ln (J))^{-1}$ in $d=4$. 
\end{abstract}

\pagestyle{plain} \setcounter{page}{1}

\maketitle

\section{Introduction}

The $O(N)$  scalar theory  is the simplest field theory with a continuous global symmetry, hence it has been 
widely used as a test to check new methods and analytic or numerical
approximation schemes and its properties have been studied in detail. 
A very important property that is realized  in this theory is the spontaneous symmetry breaking of the
global symmetry, a phenomenon that plays a central role 
both in quantum field theory and in  the study of  phase transitions in 
statistical physics. 
In particular, the $O(N)$ theory is  the simplest model suitable 
for studying the universal properties of the QCD chiral phase transition at finite 
temperature, as well as other aspects of the chirally broken phase of quark models. 

\step

The spontaneous breaking of a global symmetry is signaled by a nonzero value of the 
order parameter,
which in the $O(N)$ theory corresponds to a nonvanishing vacuum expectation value of 
one component  of the scalar field, and  is typically associated with the 
appearance of massless excitations, the Goldstone bosons\cite{goldstone},  whose propagation induce 
infrared divergences that can substantially modify the mean field indications.
The nature and the strength of the infrared divergences is strictly related to the number of  Euclidean 
dimensions $d$ of the system and in particular the Coleman-Mermin-Wagner
theorem, specifically formulated for spin models in\cite{mermin} and for quantum fields in \cite{coleman}, 
precludes the existence of a phase with conventional long range order, or spontaneous symmetry breaking 
in $d\leq2$. When $d$ increases, the infrared divergences become less severe
and one has to resort to a suitable approximation scheme in order to deal with the infrared 
sector and extract finite answers for the physical observables.

\step

A specific observable of the $O(N)$ theory  where the effect of the Goldstone bosons is crucial, is the 
mass of the  field that acquires a nonvanishing vacuum expectation value $\varphi_0$, indicated as the longitudinal field. 
In fact, while  it is possible to make use of general symmetry arguments to put constraints on the mass 
of the Goldstone bosons associated to the other $(N-1)$ transverse fields, the same is not true for 
the longitudinal field. When the scalar theory is coupled  to  a magnetic field $J$,  the Ward-Takahashi
identities, which relate the various correlation functions of the theory according to the symmetry properties
of the problem, state that the inverse  transverse susceptibility, defined as the inverse transverse propagator
at zero momentum, is $\chi^{-1}_T=\Gamma_T(p=0)=J/\varphi_0$. Then, when the magnetic field $J$ is turned off the  
Ward-Takahashi identities ensure the vanishing of $\Gamma_T(p=0)$, that is a vanishing Goldstone bosons mass \cite{zinn}.

\step

There is no such constraint on the inverse longitudinal  susceptibility $\chi^{-1}_L=\Gamma_L(p=0)$
and therefore on the mass of the longitudinal field. However, the development of Renormalization Group 
techniques provided a deeper understanding of the infrared structure of the theory,  and
for this specific problem  in $2<d<4$, the relation  $\chi^{-1}_L\propto J^{(4-d)/2}$ is predicted
\cite{brezin1,brezin2,lawrie}.
This scaling of the longitudinal mass has been repeatedly analyzed and confirmed in various approaches
\cite{Anishetty:1995kj,Pelissetto,Engels:1999dv,Pelissetto:2000ek,ParisenToldin:2003hq}.
However the behavior of this quantity in $d=4$ is different  because in this 
case the power law scaling is replaced by much weaker logarithmic corrections. 
In \cite{Anishetty:1995kj}
a few arguments are presented to show that $\chi^{-1}_L$ also vanishes in $d=4$ and
in \cite{Zappala:2012wh}, where the two point function of the $O(N)$ theory is studied in the framework of 
the Functional Renormalization Group,  
a numerical investigation suggests a vanishing mass in $d=4$.

\step

The Functional Renormalization Group is a reformulation of the Wilsonian Renormalization Group
\cite{Wilson:1973jj}, 
based on the infinitesimal integration of momentum modes from a path integral representation
of the theory with the help of a Wilsonian momentum cutoff.  The resulting functional flow equations interpolate 
between the  microscopic theory at short distances and the full quantum effective theory at large distances.
Various realizations of the Functional Renormalization  Group were developed and, among the most renowned, there 
are Polchinski's and  Wetterich's  formulations 
\cite{Polchinski:1983gv,Wetterich:1992yh} (for reviews see \cite{Bagnuls:2000ae} \cite{Berges:2000ew}).
Although this method is  particularly suitable to analyze the critical domain 
and  is especially powerful in the evaluation of the related  critical exponents (see for 
instance  \cite{Mazza:2001bp,Litim:2010tt,Benitez:2011xx} for an extended list of references), 
it also turns out to be  an 
important tool in the study  of nonperturbative 
features of the effective potential and effective  action.  For instance, the effective potential of the broken 
phase of scalar theories with its convex form (this convexity property, \cite{syma,ilio,curt}, 
does not show up 
in perturbation theory\cite{rivers} and can only be recovered in some nonperturbative scheme \cite{convexity})
has been largely investigated by means of Renormalization Group techniques
\cite{ringwald,tetradis1,aoki,alexander,tetradis2,d1,Andronico:2002bb,Branchina:2003kf,
bonannolac,Litim:2006nn,consolizap,caillol},\cite{Zappala:2012wh}. 
In particular,
an  improvement on the 
determination of the effective potential in the ordered phase of the  $O(N)$ theory is considered in 
\cite{Zappala:2012wh}, 
where a numerical computation of the longitudinal propagator which  includes the wave function renormalization, 
i.e. momentum dependent corrections to the Local Potential approximation, is performed in $d=4$ with the result that large 
wave function renormalization factors appear at least  away from the critical  region of the phase transition to 
the disordered phase. 

\step

In this paper we reconsider the study of the longitudinal mass in the broken phase of the $O(N)$ theory, in the framework 
of the Functional Renormalization Group, generalizing the case $d=4$ examined in \cite{Zappala:2012wh} 
to the range $2<d\leq 4$, however not including momentum dependent corrections that  are not 
crucial for our purpose and therefore limiting ourselves to the Local Potential
approximation.
We first analyze the problem in the simpler  large $N$ limit case and compare 
the results with those obtained at lowest order in  the $1/N$ expansion.  Then we  extend our investigation 
to small values of $N$, also confronting  the results  with the 
Renormalization Group improved effective potential 
as obtained from the Callan-Symanzik equations. 
With this compared study,  we gain 
analytic control on the problem which allows to overcome the limits of 
the numerical analysis, that are particularly evident  in proximity of $d=4$,
and finally we get  a more complete picture of this specific issue. 
In Sec. \ref{largen} we briefly go through the  determination of the longitudinal susceptibility at 
the leading order in the $1/N$ expansion and in  Sec. \ref{RGFLOW} we discuss the Functional Renormalization 
Group approach in the same limit. In  Sec. \ref{finiten}   the problem at small $N$ is investigated by means of the 
Functional Renormalization 
Group techniques and of the perturbative Renormalization Group 
improvement.  
Conclusions are reported in  Sec. \ref{discussion}.

\section{The large $N$ limit}
\label{largen}

We start by considering the $O(N)$ theory defined by the partition function
\begin{eqnarray}
\label{zetajninf}
 Z[J]=\int \mathcal{D}{\Phi}\, e^{-S[\Phi]+\int J\cdot \Phi}
\end{eqnarray}
where $\Phi$ is a $N$-component scalar field and the action $S[\Phi]$ reads
\begin{eqnarray}
\label{classact}
 S[\Phi]=\int d^d x\,\left \{\frac{1}{2}\left [ \partial_\mu \Phi\right ]^2+\frac{M^2}{2}
\Phi^2+\frac{\lambda}{4!}\left [\Phi^2 \right ]^2 \right\} \, ,
\end{eqnarray}
and we focus on  the limit of large number of field components, $N\to \infty$.
By using the standard technique of introducing an auxiliary field $\rho$
in the partition function via the following constant multiplicative factor
(where the integration contour in the complex $\rho$ plane must be taken parallel to the imaginary axis)
\begin{eqnarray}
\label{auxiliary}
\int \mathcal{D}\rho \, e^{\frac{3}{2\lambda} \int d^dx\left[\rho - \left(M^2 +
\frac{\lambda}{6}\Phi^2 \right) \right ]^2 }=
\int \mathcal{D}\rho \,
e^{\int d^dx\left[\frac{3}{2\lambda}\rho^2-\frac{3 M^2}{\lambda}\rho-\frac{1}{2}\rho\Phi^2
+\frac{M^2}{2}\Phi^2 +\frac{\lambda}{4!}\left(\Phi^2\right)^2
+\frac{3 M^4}{2\lambda}
\right]}
\end{eqnarray}
and then, by indicating with $\varphi$ the longitudinal field component parallel to the source $J$
and integrating over the remaining $(N-1)$ components of $\Phi$,
the new effective action, functional of the fields $\varphi$ and $\rho$, that appears in the
partition function
\begin{eqnarray}
Z[J]=\int \mathcal{D}\varphi \mathcal{D}\rho \,e^{-S_{eff}[\varphi,\rho]
-\int d^d x\,(J,\varphi)}
\end{eqnarray}
is:
\begin{eqnarray}
\label{effactninf}
&&S_{eff}[\varphi,\rho] = 
\int d^dx\, \left [ \frac{1}{2}(\partial_{\mu}\varphi)^2+
\frac{1}{2}\rho\varphi^2+
\frac{3M^2}{\lambda}\rho
-\frac{3}{2\lambda}\rho^2  \right ]  \nonumber\\
&&+\frac{ N- 1}{2} {\rm tr} \,{\rm ln} \left [ \left(-\partial^2+\rho\right)\delta^d(x-y) \right ]
\end{eqnarray}
where we also included the coupling to the source $J$.

\step

This procedure shows the explicit dependence on $N$ and it is understood that, for large $N$,
the square field $\varphi^2$, the inverse coupling $1/\lambda$ and the effective action $S_{eff}$
are proportional to $N$.  The  limit $N\to \infty$  is typically studied by keeping the product 
$(N \,\lambda)$ finite and performing the functional integral by means of the steepest descent method.
In particular, by restricting to uniform configurations for $\varphi$ and $\rho$,  one gets
the two coupled  extremum  equations:
 \begin{eqnarray}
\frac{\delta S_{eff}}{\delta \varphi}&=&\rho\varphi= J,
\label{saddle1}\\
\frac{\delta S_{eff}}{\delta \rho}&=&-\frac{3}{\lambda}(\rho-M^2)+\frac{\varphi^2}{2}+\frac{N}{2}
\int \frac{d^d p}{(2\pi)^d}\frac{1}{p^2+\rho}=0
\label{saddle2}
\end{eqnarray}
and  the fluctuations around the solutions of
Eqs.(\ref{saddle1},\ref{saddle2}) are neglected because they are suppressed by inverse
powers of $N$.

\step

 From Eqs.(\ref{saddle1},\ref{saddle2}) at $J=0$ one can
immediately find the symmetric phase solution where $\varphi=0$ and
$\rho$ is the selfconsistent square mass solution of the gap
equation (\ref{saddle2}). Clearly this is possible only if $M^2 >
M^2_c$, where
\begin{eqnarray}
\label{mcritninf}
 M^2_c=-\frac{\lambda N}{6}\int \frac{d^d p}{(2\pi)^d}\frac{1}{p^2}.
\end{eqnarray}
For $M^2 < M^2_c$ the system admits another solution and the broken phase
is realized.

\step

The broken phase, which will be considered below, corresponds
to a nonvanishing field $\varphi\neq 0$ at $J=0$. This, according to
Eq. (\ref{saddle1}) , implies $\rho=0$ and  at the same time Eq.
(\ref{saddle2}) yields the value of the field $ \varphi_0$ that correponds 
to the minimum of the potential 
\begin{eqnarray}
\label{phi0ninf} \varphi_0^2=-\frac{6 M^2}{\lambda}+\frac{6
M_c^2}{\lambda}.
\end{eqnarray}
Then we are able to rephrase Eqs.(\ref{saddle1},\ref{saddle2}) at
finite $J$, in terms of $\varphi_0$
\begin{eqnarray}
\rho&=& \frac{J}{\varphi_0} \left( 1  + \frac{ \varphi -\varphi_0}{\varphi_0} \right )^{-1} \,
\label{sadbrok1}\\
\varphi^2 -\varphi_0^2 &=& \frac{6}{\lambda}\rho - N \int \frac{d^d
p}{(2\pi)^d} \left ( \frac{1}{p^2+\rho} - \frac{1}{p^2} \right )
\label{sadbrok2}
\end{eqnarray}
and it is easy to check  the dependence of the field
$\varphi$ on the source $J$, at least in $2< d <4$, where the integral 
in the right hand side of Eq. (\ref{sadbrok2}) is finite and proportional
to $\rho^{\frac{d-2}{2}}$. After performing  the integral, 
Eq. (\ref{sadbrok2}) becomes
\begin{equation}
\label{sadagain}
\varphi^2 -\varphi_0^2 =
\frac{6 \rho}{\lambda} 
- N \frac{\Gamma(1-d/2)}{(4\pi)^{d/2}}\rho|\rho|^{\frac{d-4}{2}}.
\end{equation}
Eq.\,(\ref{sadagain}) can be explicitly solved for $\rho$, at least in the region $ | \rho | <<1$, 
(which means close to the minimum of the potential, according to the saddle point
equations (\ref{saddle1},\ref{saddle2}) with $\varphi\neq 0$ )
where the term $\rho^{\frac{d-2}{2}}$ is dominant with respect to the linear term $\rho$, 
\begin{equation}
\label{ro}
\rho = \left [\theta(\varphi^2+\varphi^2_0) - \theta(\varphi^2_0-\varphi^2) \right ]
\left(-(4\pi)^{d/2}\frac{|\varphi^2 -\varphi^2_0|}{N\Gamma(1-d/2)}\right)^{\frac{2}{d -2}}.
\end{equation}
The solution (\ref{ro}), when replaced in Eq. (\ref{saddle1}), 
$V'(\varphi)=\varphi \rho$, yields an analytic expression of the derivative of the potential 
in the region  $ | \rho | <<1$.

\step 

Incidentally  when  $M^2=M^2_c$,
i.e. $\varphi_0=0$, one finds:
\begin{equation}
\label{critdependence} \varphi \propto J^{\frac {d-2}{d+2}} 
\end{equation}
and the inverse of the exponent of $J$ gives the correct 
critical index $\delta=(d+2)/(d-2)$.  
This picture is slightly modified in the
broken phase with $\varphi_0 \neq 0$, again for $2< d <4$. 
In fact in this case it is
convenient to define the shifted field $\varphi_{sh}= \varphi
-\varphi_0$ so that in the left hand side of Eq. (\ref{sadbrok2}),
for $J\to 0$ one can retain the $O(\varphi_{sh})$ term, 
neglecting  $O(\varphi^2_{sh})$ and, in the same way, the leading behavior of $\rho$
is given by $\rho \sim J/\varphi_0$. Then, as before, in the right hand
side of  Eq. (\ref{sadbrok2}) one can retain only the leading term
proportional to $\rho^{\frac{d-2}{2}}$ and finally one finds
\begin{equation}
\label{brokdependence} \varphi_{sh} \propto J^{\frac {d-2}{2}}
 \, .
\end{equation}
It is then straightforward to get the behavior of 
$(\partial \varphi (J) / \partial J)$ 
from the derivative of Eq. (\ref{brokdependence}) 
for small $J$ and $2<d<4$, by recalling that $\varphi_0$ in Eq. (\ref{phi0ninf})
is independent of $J$,
\begin{equation}
\label{suscdependence} 
\chi_L^{-1}(J)=\left ( \frac{\partial \varphi}{\partial J}
\right )^{-1} \propto J^{\frac {4-d}{2}}
 \, .
\end{equation}
Then, clearly, for  $2<d<4$ when the source $J$ is turned off 
we have $\chi_L^{-1}(J=0)=0$.

\step

In $d=4$, the integral in Eq. (\ref{sadbrok2}) contains a
logarithmic divergence and the expression in Eq. (\ref{brokdependence}) is modified by
logarithmic corrections so that it is no
longer straightforward to generalize Eq. (\ref{suscdependence})
to $d=4$ by following the same procedure.
Therefore, we shall reconsider the problem by analyzing in detail the
behavior of the integral, including the case $d=4$. In particular we
use Eqs.(\ref{sadbrok1},\ref{sadbrok2}) to express
$\rho=\rho(\varphi,J)$ and finally  $\varphi$ as a function of the
source $J$:  $\varphi=\varphi(J)$. Therefore, by deriving
Eqs.(\ref{sadbrok1},\ref{sadbrok2}) with respect to $J$ and putting
them together one has
\begin{equation}
\label{susc}  \chi_L^{-1}(J)=\left ( \frac{\partial \varphi (J)}{\partial J} \right )^{-1} 
= \rho +\frac{2\varphi^2}{\frac{6}{\lambda}+ N \, I^{(d)}_2(\rho)} 
\end{equation}
where
\begin{equation}
\label{integral}
I^{(d)}_2(\rho) = \int\frac{d^dp}{(2\pi)^d}
\frac{1}{(p^2+\rho)^2}
\end{equation}

\step

The inverse longitudinal susceptibility is given by  Eq. (\ref{susc}) in
the limit $J\to 0$. Incidentally we note that the same result as in
Eq. (\ref{susc}) is obtained as well,  when 
computing the longitudinal mass $ m_L^2(J)=\chi_L^{-1}(J)$ from the
inverse propagator of the theory $\delta S_{eff} / ( \delta \varphi
\delta \varphi )$ (by treating $\rho$ as $\rho(\varphi)$ ) at zero
momentum. To extract the correct behavior of Eq. (\ref{susc}) when
$J\to 0$ it is essential to analyze in detail the integral  $I^{(d)}_2(\rho)$ 
including the case $d=4$. We compute $I^{(d)}_2(\rho)$ by inserting an ultraviolet 
Euclidean cutoff $\Lambda$ which protects the result from possible ultraviolet divergences:
\begin{eqnarray}
\label{idue}
 I^{(d)}_2(\rho)&=& C_d |\rho|^{\frac{d-4}{2}}
 \int^{\Lambda^2/|\rho|}_0 dx \, \frac{x^{\frac{d-2}{2}}}{(x+1)^2}\nonumber\\
 &=&C_d|\rho|^{\frac{d-4}{2}}\left[\frac{1}{1+\frac{\Lambda^2}{|\rho|}}+\frac{(d-2)}{d}
 \phantom{a}_2F_1\left(1,\frac{d}{2};\frac{d+2}{2};-\frac{\Lambda^2}{|\rho|}\right)\right]\left(\frac{\Lambda^2}{|\rho|}\right)^{d/2}
 \nonumber\\
 &=&C_d|\rho|^{\frac{d-4}{2}}\left[-\frac{d-2}{d}\Gamma(1-d/2)\Gamma(1+d/2)
 +\frac{2}{d-4}\left(\frac{|\rho|}{\Lambda^2}\right)^{\frac{4-d}{2}}\right.\nonumber\\
 &-&\left.\frac{4d}{(d-6)}\left(\frac{|\rho|}{\Lambda^2}\right)^{\frac{6-d}{2}}
 +\mathcal{O}\left[\left(\frac{|\rho|}{\Lambda^2}\right)^{\frac{8-d}{2}}\right]\right]\nonumber\\
 &=&|\rho|^{\frac{d-4}{2}}\left(\frac{\Gamma(2-d/2)}{(4\pi)^{d/2}}
 +\frac{2}{(d-4)(4\pi)^{d/2}\Gamma(d/2)}\left(\frac{|\rho|}{\Lambda^2}\right)^{\frac{4-d}{2}}+\dots\right)
\end{eqnarray}
where $C_d \equiv ( (4\pi)^{d/2}\Gamma(d/2) )^{-1}$, $\Gamma(x)$ is the Gamma function and 
$\phantom{a}_2F_1 \left(a,b;c;z \right)$ is the Gaussian Hypergeometric function.

\step

We note that Eq. (\ref{idue}) when $d<4$ does not present any ultraviolet divergences.
The leading term is $ [ \Gamma(2-d/2) / (4\pi)^{d/2} ] |\rho|^{\frac{d-4}{2}}$ and all
the other terms in the development of Eq.\,(\ref{idue}) are suppressed by positive 
powers of $|\rho| / \Lambda^2$.
If we take $\Lambda\to \infty$, only the leading term survives and we end up with the same result already 
shown in Eq. (\ref{suscdependence}). However, due to the presence of the cutoff
$\Lambda$,  we are able to analyze the case $d=4$. In fact one can easily check that,
by expanding the last line of  Eq. (\ref{idue})  in powers of $\epsilon=4-d$, in the limit  $d\to 4$
the divergent terms in $\epsilon$ cancel out leaving a logarithmic term  :
\begin{equation}
\label{iduequattro}
I^{(4)}_2(\rho) = -\frac{1}{16\pi^2}\left(\ln\frac{|\rho|}{\Lambda^2}+1\right)
\end{equation}
In this case the ultraviolet cutoff $\Lambda$ appears logarithmically and
we have to recall that for $d=4$ the limit of infinite $\Lambda$ corresponds to a trivial 
behavior of the theory that turns out to be noninteracting at any finite renormalization scale.
Therefore we limit ourselves to the study  of an effective theory with fixed ultraviolet cutoff $\Lambda$.
Then the linear term in $\rho$ in  Eq. (\ref{susc}) is negligible and the inverse susceptibility  goes to zero as  
$(|\ln (J)|)^{-1}$ , when $J\to 0$.  We shall analyze in more detail this aspect in Sec. \ref{finiten}.

\section{Functional Renormalization Group  at large $N$}
\label{RGFLOW}

In this Section we shall use the Functional Renormalization
Group flow equations  to analyze the theory in the large $N$ limit.
According to this approach one can follow the evolution of the running 
effective action $\Gamma_k$ as a function of a momentum $k$ scale which,
starting from the ultraviolet cutoff $\Lambda$, goes down  to 
the infrared limit $k=0$.  The flow of $\Gamma_k$  
starts in the ultraviolet region with an initial boundary condition 
given by the bare action and its final endpoint at $k=0$ corresponds 
to the full effective action of the theory. The integration of the fluctuations
down to the scale $k$, which generates the flow of $\Gamma_k$ , is 
performed by modifying the action of the theory with an additional term, quadratic 
in the fieds, that contains a $k$-dependent regulator $R_k$ which acts like an
infrared cutoff that effectively enables the integration of the modes above $k$. In order to 
fullfill the requirements that $\Gamma_{k=\Lambda}$ is the bare action 
and $\Gamma_{k=0}$ coincides with the full effective action, the regulator $R_k$ 
must have the following asymptotic properties : $R_{k\to \infty} =\infty$ and
$R_{k=0} =0$. Then if $\Phi=\left(\Phi_1,\dots,\Phi_N\right)$,
$\Gamma^{(2)}_k\left(q,-q;\Phi\right)_{a,b}$ is the second functional derivative:
\begin{eqnarray}
\Gamma^{(2)}_k\left(q,-q;\Phi\right)_{a,b}
=\frac{\delta^2\Gamma_k[\Phi]}{\delta\Phi_a(q)\delta\Phi_b(-q)},
\end{eqnarray}
$[R_k(q)]_{ab}$ is the $k$-dependent regulator and
$t$ is defined as $t=\ln (k/\Lambda)$,  the flow equation reads:
\beq \label{rgfloweq}
\partial_t \Gamma_k[\Phi]=\frac{1}{2} \int \frac{d^dq}{(2\pi)^d}\, \Tr
\left(\partial_t R_k(q)
\left [ \Gamma_k^{(2)}[q,-q;\Phi]+R_k(q)\right ]^{-1}\right)
\eeq
where the trace is performed over the internal indices.

\step 

In order to analyze the behavior of the longitudinal susceptibility for the $O(N)$ theory,
it is sufficient to
approximate Eq. (\ref{rgfloweq}) to the so called 
Local Potential approximation,
where we limit the form of $\Gamma_k[\Phi]$ by  ignoring all possible scale dependent 
coefficients of the kinetic $\left ( O(\partial^2) \right )$ term  
and by fully neglecting $O(\partial^4)$ terms,
\begin{equation}
\Gamma_k[\Phi]=\int_x\left(  U_k(|\Phi|)  +\frac{1}{2}
\left[ \partial_\mu \Phi_a \partial_\mu \Phi_a \right ] + O(\partial^4) \right) \, .
\label{dexpansion}
\end{equation}
In Eq. (\ref{dexpansion}), due to the symmetry of the problem, the potential $U_k$ depends on the modulus $|\Phi|= \sqrt{\Phi_a\Phi_a}$
and, according to our approximation, Eq. (\ref{rgfloweq}) reduces to a flow equation 
for the potential $U_k(|\Phi|)$,  with ultraviolet boundary conditions  fixed in Eq. (\ref{classact}) :
\begin{equation}
\label{potbound}
U_{k=\Lambda}(|\Phi|) = \frac{M^2}{2} \Phi^2+\frac{\lambda}{4!}\left [\Phi^2 \right ]^2 
\end{equation}
and the final infrared limit of the  flow,  $U_{k=0}(|\Phi|)$, corresponds to 
(an approximation of ) the full effective potential.

\step 

The explicit form of the flow equation depends on the particular choice of
the matrix $R_k$,
and here it is convenient to take the following regulator that allows us to 
solve analytically the momentum integrals  \cite{Litim:2000ci,Litim:2001up,Litim:2001fd,Pawlowski:2005xe}:
\beq
\label{optimctf}
[R_k(q)]_{ab} =(k^2 -q^2) \;\;\Theta (k^2 -q^2)\,\delta_{ab}
\eeq
where $\Theta(x)$ indicates the Heaviside function. With the regulator in Eq. (\ref{optimctf}),
the flow equation for $U_k(|\Phi|)$ is 
\begin{equation}
\label{simplerflow}
\partial_t U_k(|\Phi|) =\frac{2 C_d}{d} k^{2+d} \left [  \frac{(N-1)}{k^2 +  U'_k(|\Phi|)/ |\Phi| }
+ \frac{1}{k^2 + U''_k(|\Phi|)} \right ]\; ,
\end{equation}
where $C_d$ is defined after Eq. (\ref{idue}) and the 
primes indicate derivatives with respect to $|\Phi|$. The contribution
of the $(N-1)$ Goldstone bosons and that of the  longitudinal field in Eq. (\ref{simplerflow})
are clearly evident.

\step

The very simple structure of Eq. (\ref{simplerflow}) allows us to analyze the large $N$ limit,
where the contribution of the longitudinal component is negligible and therefore, by deriving 
with respect to $|\Phi|$, we get 
\begin{equation}
\label{simplerflow2}
\partial_k U'_k(|\Phi|) =  \frac{2 C_d}{d}  k^{1+d}\; \frac{N}{|\Phi|}\;\;
\frac{   U'_k(|\Phi|)/ |\Phi| - U''_k(|\Phi|) }{\left(k^2 +  U'_k(|\Phi|)/|\Phi| \right)^2} \, .
\end{equation}
We recall that the inverse longitudinal susceptibility coincides with the square mass of the longitudinal 
field which is the second derivative of the effective potential evaluated at the vacuum expectation
value of the field, i.e. at the minimum of the effective potential.  
Therefore in this framework  it corresponds to  $U''_{k=0}(|\Phi_0|)$ with $|\Phi_0|$ such that
 $U'_{k=0}(|\Phi_0|)=0$ . The flow equation in the form of Eq. (\ref{simplerflow2}) allows us to
make some statements on  $U''_{k=0}(|\Phi_0|)$. 

\step 

Before doing that, we have first to recall the behavior of the running potential in the broken phase 
when $k\simeq 0$, which  has repeatedly been described in various contexts
\cite{aoki},\cite{tetradis2},\cite{alexander},\cite{d1},\cite{bonannolac,Litim:2006nn,consolizap,caillol},\cite{Zappala:2012wh}.
Essentially, it turns out that for  $k$ much smaller than the other characteristic scales of the problem, 
the derivative of the potential is very well approximated by 
$U'_{k}(|\Phi|)= - k^2 |\Phi| + O(k^3, |\Phi|^2) $ in the region  $0<|\Phi| \lesssim |\Phi_0|$, 
while for  $|\Phi|\gtrsim |\Phi_0|$, $U'_{k}(|\Phi|)$ has already reached its infrared 
limit and is practically  $k$ - independent with the exception of a very narrow region around 
$|\Phi_0|$  where the two different branches merge. 
In this narrow region, no discontinuity is observed in $U'_{k}(|\Phi|)$, and in particular one has 
$U'_{k=0}(|\Phi_0|)=0$. 
\begin{figure}[h]
\begin{minipage}[c]{0.49\textwidth}
    \epsfxsize=8cm
    \epsfysize=6.2cm
    \centerline{\epsffile{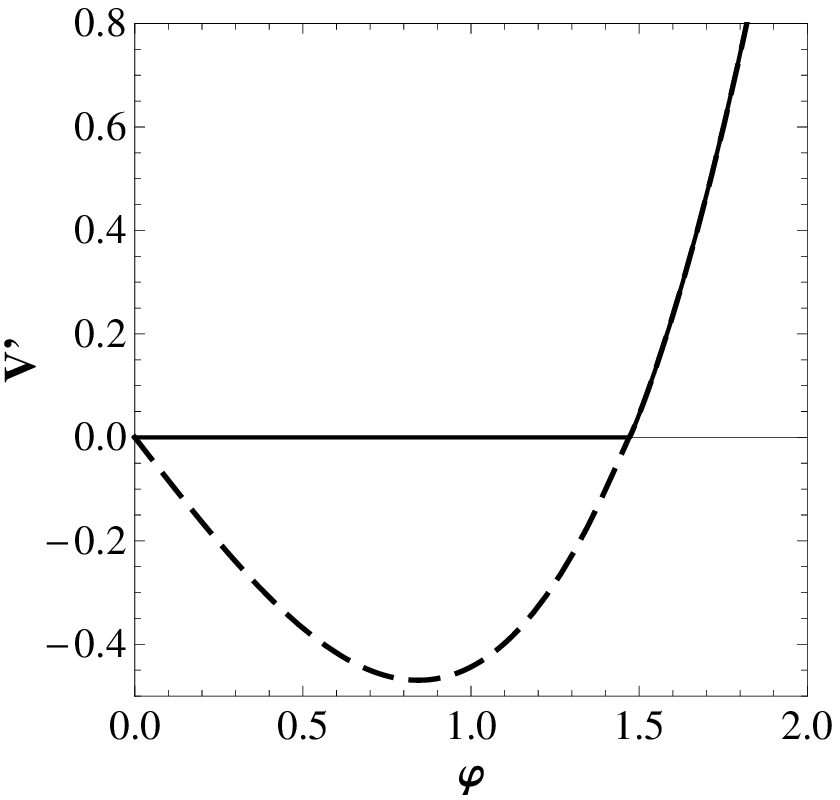}}
\end{minipage}
\begin{minipage}[c]{0.49\textwidth}
    \epsfxsize=8cm
    \epsfysize=6cm
    \centerline{\epsffile{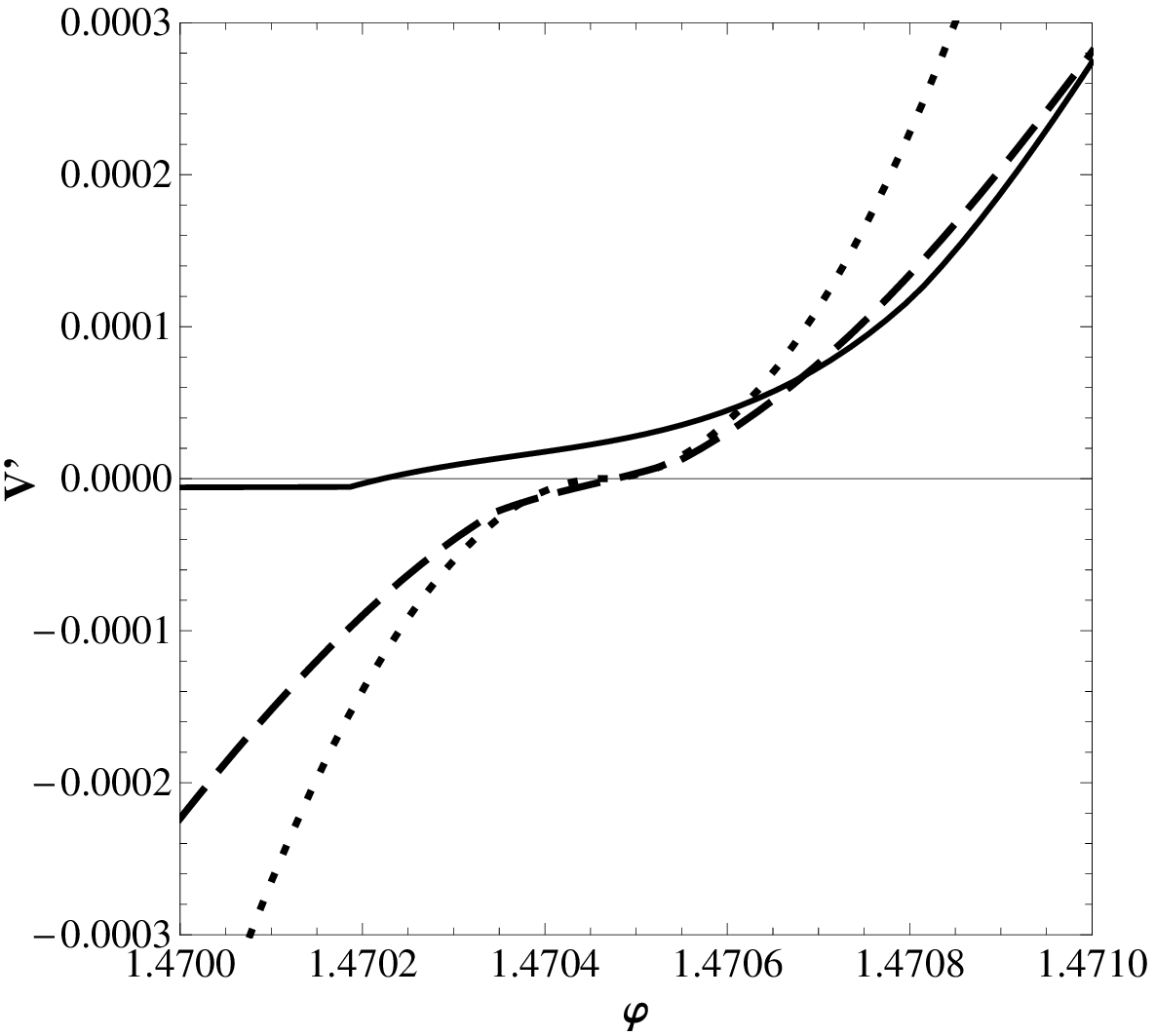}}
\end{minipage}
\caption{\label{figra1}
Left Frame. The derivative of the effective potential in the large $N$ limit given in Eqs. (\ref{saddle1}) and (\ref{saddle2})
(dashed) compared with the Functional Renormalization Group $U'_k(|\Phi|/\sqrt{N})/ \sqrt{N}$ at $k=0.002$ (solid)
for $d=3$,  $M^2 =-1$, $N \lambda=2.4$, $\Lambda=10$.\\
Right Frame. Zoom of the Left Frame in the region around $\varphi_0=1.47046$ , with the inclusion of the analytic solution 
in Eq. (\ref{ro}) (dotted).
}
\end{figure}
Accordingly, one finds that $U''_{k}(|\Phi|) \sim k^2 $ 
for  $0<|\Phi| \lesssim |\Phi_0|$ and therefore, when  $k\to 0$, its left limit 
(when  $ |\Phi|-|\Phi_0|\to 0^-$) vanishes.
Our goal is to establish  whether or not  its right limit (when $ |\Phi|-|\Phi_0|\to 0^+$)
vanishes, i.e. whether or not $U''_{k=0}(|\Phi_0|)$ is discontinuous.

\step

In order to clarify this point, we observe that
according to the mentioned  solution for $U'_{k}(|\Phi|)$ at $k\simeq 0$,
the denominator in Eq. (\ref{simplerflow2}),
computed at $|\Phi_0|$, vanishes as $k^6$ and, when combined with the other powers of 
$k$ in Eq. (\ref{simplerflow2}) , it gives the factor $k^{d -5}$ which 
diverges in the infrared limit $k \to 0$ (in our problem $d\leq 4$).
As the left hand side of Eq. (\ref{simplerflow2}) at $|\Phi_0|$ and for $k\to 0$  cannot be divergent,
the numerator in the right hand side must vanish rapidly enough to compensate the diverging factor.  
Since the continuous function  $U'_{k}(|\Phi_0|)$ vanishes as $k^2$, it follows that  
also $U''_{k}(|\Phi_0|)$ must vanish when $k\to 0$  and therefore no discontinuity 
can appear  because its presence  would produce a divergence 
in the right hand side of the differential equation when $k\to 0$. 
This analysis indicates that in the large $N$ limit the flow equations constrain the 
inverse susceptibility to vanish as   $U''_{k}(|\Phi_0|)= - k^2 + O(k^3)$. 
Unfortunately this argument cannot be straightforwardly extended to the case of 
finite $N$, because of the presence of the longitudinal 
contribution in the differential flow equation.

\step

\begin{figure}[h]
\begin{minipage}[c]{0.49\textwidth}
    \epsfxsize=8cm
    \epsfysize=6.2cm
    \centerline{\epsffile{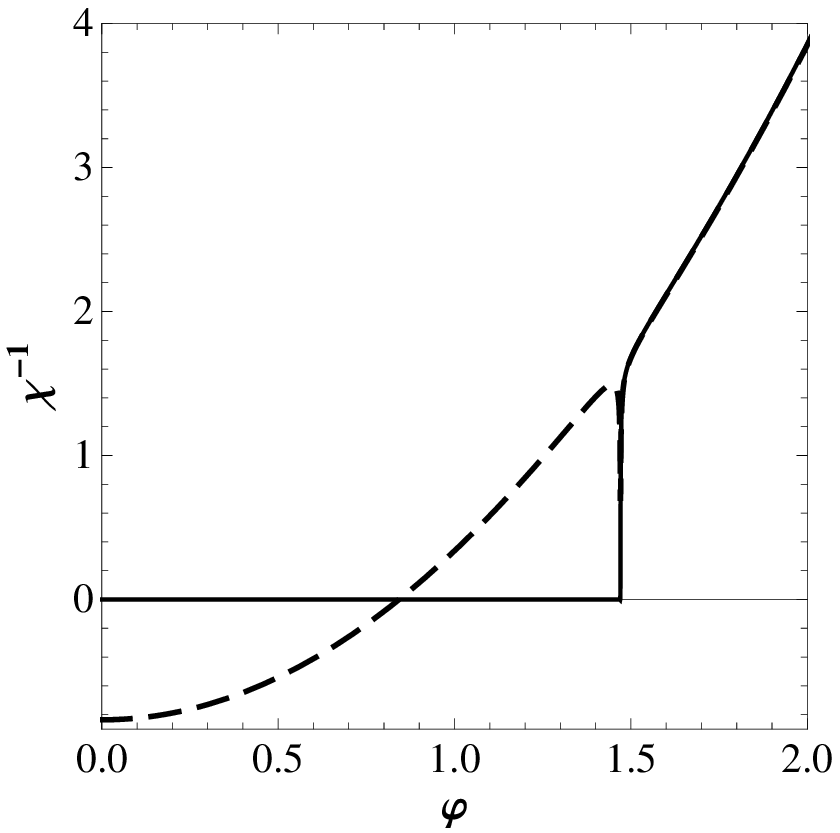}}
\end{minipage}
\begin{minipage}[c]{0.49\textwidth}
    \epsfxsize=8cm
    \epsfysize=6cm
    \centerline{\epsffile{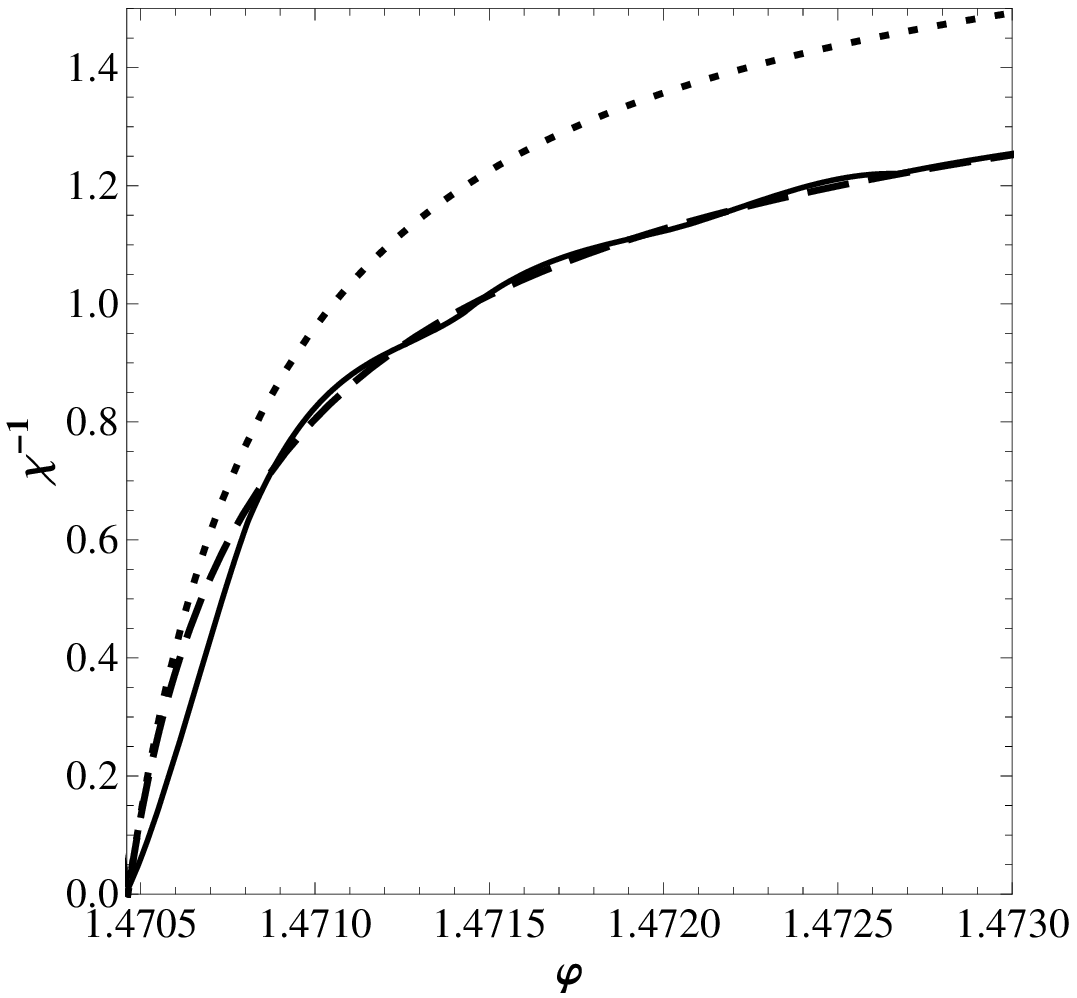}}
\end{minipage}
\caption{\label{figra2} Left Frame. The inverse longitudinal susceptibility 
in the large $N$ limit given in Eqs. (\ref{susc}) and (\ref{saddle2})
(dashed) compared with the Functional Renormalization Group $U''_k(|\Phi|/\sqrt{N})$ at $k=0.002$ (solid)
with the same parameters used in Fig. \ref{figra1}.\\
Right Frame. Zoom of the Left Frame in  the region around $\varphi_0=1.47046$, with the inclusion of the analytic solution in Eq. (\ref{ro})
(dotted).  
}
\end{figure}

On the other hand, one can try to extract the longitudinal
susceptibility by numerically solving the Functional Renormalization
Group flow. In fact the numerical resolution of the flow consists in
the integration of a partial differential equation for the potential
$U_k(|\Phi|)$ with fixed ultraviolet boundary conditions indicated
in Eq. (\ref{potbound}). The integration is carried out with the infrared
scale that approaches $k=0$ and the corresponding  output of the
flow provides a numerical approximation of the effective potential.
Further details of the numerical integration can be found in
\cite{Zappala:2012wh}.

\step

The numerical output of the Functional Renormalization Group flow
can be compared with the solution of the saddle point equations
obtained in the large $N$ limit. In particular, we can compare $U'_k$
and $U''_k$ with the first and second derivative of the potential
obtained from  Eqs. (\ref{saddle1},\ref{susc}) where the parameter $\rho$ is
replaced according to Eq. (\ref{saddle2}). We can also check the
analytic solution for $\rho$ displayed in Eq. (\ref{ro}), which is
 valid only in a narrow region around the minimum. As a specific example,
in Figs. \ref{figra1} and \ref{figra2} the first and second
derivative of the potential are displayed for  $d=3$ and  with the following  
set of bare parameters, $M^2 =-1$, $N \lambda=2.4$. The ultraviolet cutoff is taken
$\Lambda=10$ for the Functional Renormalization Group and it is
rescaled of a factor $2/3$ in Eqs. (\ref{saddle1},\ref{saddle2},\ref{susc}), 
because of  the difference in the dependence of the renormalized 
parameters on $\Lambda$, as discussed in \cite{Zappala:2012wh}.

\step

\begin{figure}
${}$\vskip1cm \epsfig{file=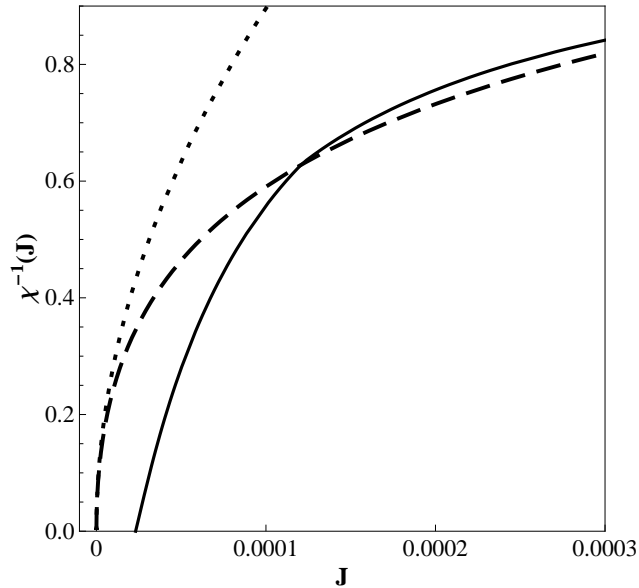,width=.5\hsize}
\caption{\label{figra3}
The inverse longitudinal susceptibility 
in the large $N$ limit plotted as a function of the source $J$.
All parameters are fixed as in Figs. \ref{figra1} and \ref{figra2}.
The same coding of Fig. \ref{figra2} is used for dotted, dashed and solid curves.
}
\end{figure}

In the limit $k\to 0$, the Functional Renormalization Group flow
produces a convex potential  with a flat part in the region
$|\Phi|\leq |\Phi_0|$ ,
\cite{ringwald,tetradis1,aoki,alexander,tetradis2,d1,Andronico:2002bb,Branchina:2003kf,
bonannolac,Litim:2006nn,consolizap,caillol},\cite{Zappala:2012wh},
in agreement with a basic property of the
effective potential \cite{syma,ilio,curt}. Actually, the numerical routine adopted here
fails to integrate the flow equations below a particular value of
the infrared scale which corresponds to $k=2 \cdot 10^{-3}$ in the example
considered, and this explains the small difference from zero
observed in the Functional Renormalization Group curve of Fig. \ref{figra1}
(right frame) for $|\Phi|\leq |\Phi_0|$.
It is also well known that the convexity property  of the potential is
not recovered when using approximations based on an expansion around
a saddle point, such as the loop expansion or the large $N$
expansion considered in Sec. \ref{largen}. 
In these cases, it is necessary to consider
the contribution of all the saddle points in order to
obtain  a convex potential \cite{rivers}. This explains why in Fig. \ref{figra1}
we observe that at the leading order in the $1/N$ expansion, despite its
claimed exactness, the effective potential does not present the expected 
convex shape in the region between $|\Phi|=0$ and $|\Phi|=|\Phi_0|$.
 
\step

On the other hand, for $|\Phi|\geq |\Phi_0|$ the computation 
performed by considering a unique saddle point
is exact in the limit $N\to \infty$ and this explains the good
agreement of the various curves  above $|\Phi_0|$. 
We observe in the right frame of Figs. \ref{figra1}  and \ref{figra2}
that the analytic solutions for $J$ (i.e. the derivative of the potential) and $\chi_L^{-1}$ 
(see Eqs.\,(\ref{saddle1},\ref{saddle2},\ref{susc},\ref{ro})), 
perfectly reproduce the numerical solution of the large $N$ equations 
only for $|\Phi|$ very close to $|\Phi_0|$, as expected.
From the plot of the curves in  Fig. \ref{figra2}
it is also evident  that the second derivative of the potential, i.e. the 
inverse susceptibility, vanishes at $|\Phi_0|$. 
Finally the plot of $U''_k$ computed via the Functional 
Renormalization Group is in good agreement with the other curves  
(compatibly with unavoidable  numerical fluctuations)
despite the flow is  truncated before reaching $k=0$ .

\step

Fig. \ref{figra3} shows the inverse suceptibility as a function of the source $J$ with  $J\geq 0$ 
for the same set  of parameters of Figs.  \ref{figra1} and  \ref{figra2}.
Again the numerical solution  of the saddle point equations (\ref{saddle1}) 
and (\ref{saddle2})  is displayed together with the analytic solution in 
Eq. (\ref{ro}) and with the output of the flow equation. Fig.  \ref{figra3}  
shows the range of small values of $J$ where the flow equation is no longer accurate. 
For larger values of $J$, the dashed and the solid lines coincide. 
In addition  we notice that the analytic solution obtained by replacing 
$\rho$ in Eq.\,(\ref{susc}) with  the result of Eq.\,(\ref{ro}), 
and which displays the  scaling behavior indicated in Eq. (\ref{suscdependence}),
does  in fact coincide with the full numerical solution only for extremely small 
values of $J$. In this  very small range, the Functional Renormalization Group 
flow equation solution has already deviated from the 
correct behavior and a more efficient algorithm becomes necessary to push 
$k$ closer to zero  to obtain a better agreement in this region
and to recover the scaling behavior of  Eq. (\ref{suscdependence}).

\begin{figure}
${}$\vskip1cm \epsfig{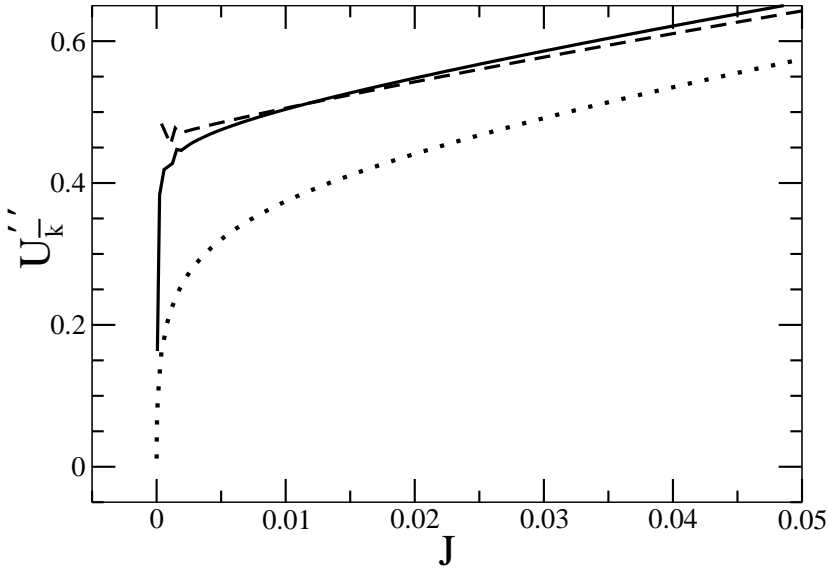}
\caption{\label{figra4}
$U''_{\overline k}$ plotted versus $J$ for $d=3$ and $\overline k=5  \cdot 10^{-4}$ (dotted) ,  $d=3.5$ and  
$\overline k= 10^{-2}$ (solid),  $d=4$ and  $\overline k= 2 \cdot 10^{-2}$ (dashed).
Values of the other parameters  : $M^2 =-1$,  $\lambda=2.4$ and $\Lambda=10$.
}
\end{figure}

\step

\section{Susceptibility  at finite N}
\label{finiten}

We start the analysis of the problem at finite $N$ by looking at the full flow equation of the potential
in Eq. (\ref{simplerflow}) with both contributions of the transverse and longitudinal fields that introduce 
the dependence of the flow on two different scales, namely the running masses of these two fields.
As already noticed it is not easy to put analytic constraints on the longitudinal mass  when $k\to 0$ 
due to the presence of these  two separate contributions. 

\step 

We turn to the numerical resolution of the flow equation to extract some physical indications
in the infrared limit. In particular, we are interested in the behavior of the inverse susceptibility,
i.e. the mass of the longitudinal field. Let us first consider the case of  
$N=4$ and bare parameters 
$M^2 =-1$, $\lambda=2.4$ and cutoff $\Lambda=10$. In $d=3$, the flow is stable down to $\overline k=5\cdot 10^{-4}$ 
and the corresponding solution $U''_{\overline k}$ is displayed in Fig.  \ref{figra4}  (dotted line) as a function 
of the source $J=U'_{\overline k}$, with  $J\geq 0$.   The other two curves in Fig.  \ref{figra4} respectively correspond 
to $d=3.5$ (solid), where the lowest reached infrared value is $\overline k= 10^{-2}$, and 
$d=4$ (dashed), with $\overline k=2\cdot 10^{-2}$ and the bare parameters have the same values as in the $d=3$ case.

\step 

Due to the finite value of $\overline k$, the three curves in Fig.  \ref{figra4} do not reach the origin $(J=0,U''=0)$. However, while for 
smaller $d$ the curves clearly approach the origin, the same is not true when $d$ is at the upper critical dimension.
In particular, when $d=3$ (dotted) the infrared cutoff $k$ can be pushed very close to $k=0$ and the corresponding curve in Fig.  \ref{figra4}
smoothly bends toward the origin. When $d=3.5$ (solid) we already observe numerical fluctuations in the plot  but even
in this case there is a change of slope and the curve approaches the origin.
In $d=4$  although the infrared cutoff  is not much larger than for $d=3.5$, 
the numerical fluctuations are stronger and make  the deep infrared region ($k\to 0$) inaccessible. 
The corresponding curve (dashed)  does not show any  bending toward the origin
and it is not possible to establish from the plot  whether or not the second derivative of 
the potential vanishes at $J=0$ and $k=0$. 
In fact,  as we shall see below by applying standard Renormalization Group  techniques,
such problem   is related to the fact that power law corrections are replaced by smoother 
logarithmic corrections in $d=4$, which turn out to be more difficult to treat numerically.
For better understanding this point, we now proceed to the study of the theory 
at finite $N$ and for $d$ close to $4$,  in the framework of  the Callan-Symanzik approach.

\step

To this end, it is convenient to rewrite the potential as
\begin{equation}
\label{potmod}
V(\phi)=\frac{\tau}{12}\phi^2+\frac{\phi^4}{24\lambda} \, ,
\end{equation}
where we have introduced the rescaled field $\phi$
\begin{equation}
\label{newfi}
\phi^2\equiv\lambda \Phi^2
\, , 
\end{equation}
and the "temperature" $\tau$ 
\begin{equation}
\label{tau}
\tau=\frac{6}{\lambda} (M^2-M_c^2)\, , 
\end{equation}
is normalized in such a way that it vanishes at the critical value $M^2=M_c^2$ that separates 
the symmetric and the broken phase.
It is understood that $M_c^2$ includes corrections with 
respect to  Eq. (\ref{mcritninf}) which is obtained in the $N\to \infty $ limit.
We are specifically interested in the broken phase and therefore
we take $\tau< 0$.

\step

We also recall that the coupling $\lambda$ has  dimension $\Lambda^{4-d}$, expressed in units of the ultraviolet cutoff $\Lambda$,
and the corresponding dimensionless coupling is defined as $u= \lambda \Lambda^{d-4}$.
By defining  $t=\ln\frac{\mu}{\Lambda}$ as the logarithm of the scale $\mu$, 
(we avoid to use the same  notation of the running momentum $k$ adopted for the 
Functional Renormalization Group flow equations due to the different roles of the two scales),  
the  $\beta$-functions of the various 
parameters in the potential, as given by perturbation theory at lowest order, are $(\epsilon =4-d ) $ :
\begin{eqnarray}
\label{beta}
\beta= \frac{du}{dt} &=& -\epsilon u+ C_d \frac{N+8}{3}u^2,\\
\label{betam}
\frac{d (M^2-M_c^2)}{dt}&=& C_d \frac{N+2}{3} u (M^2-M_c^2).
\end{eqnarray}
Note that $\Phi$ at lowest order has no correction and therefore  no $t$ dependence. 
A word of caution concerning the usual way of obtaining these $\beta$-functions
has to be said: the fact that the $O(N)$ model in the broken phase contains more than 
one scale requires a more careful treatment, \cite{Ford:1996hd},
and we shall come back again to  this point. From Eqs. (\ref{newfi},\ref{tau},\ref{beta},\ref{betam})  we get 
\begin{eqnarray}
\label{betat}
\frac{d\tau}{dt}&=&-2 C_d\tau\\
\label{betaphi}
\gamma_{\phi}&=&\frac{1}{\phi}\frac{d\phi}{dt}=
\frac{\epsilon}{2}+\frac{1}{2u}\frac{du}{dt}
=\frac{C_d}{2} \frac{N+8}{3} u  \, .
\end{eqnarray}

\step 

By solving the differential equations related to these $\beta$-functions,
one can determine the evolution of the various parameters with the scale $\mu$,
after having set the boundaries at the bare values when $\mu=\Lambda$. 
Below, the dependence of the various variables on the parameter $t$
is explicitly displayed while it is omitted for the bare  variables,
which correspond to the value $t=0$.  
So the dimensionless coupling is 
\begin{equation}
u(t)=u e^{-\epsilon t} \, \left [ 1-  \frac{C_d \, u (N+8)}{ 3 \, \epsilon } \left(1-e^{-\epsilon t} \right)
\right ]^{-1}
\label{usol}
\end{equation}
while the dimensionful coupling $\lambda(t)=u(t)\, \mu^{\epsilon}$ reads
\begin{equation}
\label{lambdasol}
\frac{1}{\lambda(t)}=\frac{1}{\lambda}+ \frac{C_d (N+8)}{ 3 \, \epsilon }
\left(\mu^{-\epsilon}-\Lambda^{-\epsilon}\right)
\end{equation}
and the temperature $\tau$ is
\begin{equation}
\label{tausol}
 \tau(t)=\tau\left(\frac{u}{u(t)}e^{-\epsilon t}\right)^{\frac{6}{N+8}} \, .
\end{equation}
Finally, the renormalization factor  $\xi(t)$ of the rescaled field $\phi$ is
\begin{equation}
\label{xisol}
 \xi(t)=e^{\int^t_0 dt'\gamma_{\phi}(t')}=
\left(\frac{u}{u(t)}e^{-\epsilon t}\right)^{-1/2}
\end{equation}

\step

We are now able to evaluate the Renormalization Group improved potential $ V_{RGI}(\phi) $, 
solution of the Callan-Symanzik equation. This is achieved by replacing
in Eq. (\ref{potmod}) the bare quantities with the corresponding  scale dependent parameters:
\begin{equation}
\label{potimpr}
 V_{RGI}(\phi)=\frac{\tau(t)}{12}\xi(t)^2\phi^2+\frac{1}{24\lambda(t)}\xi^4(t)\phi^4.
\end{equation}
The derivative of $V_{RGI}(\phi)$ (in this Section the prime 
indicates derivative with respect to $\phi$) is:
\begin{equation}
\label{dpotimpr}
V'_{RGI}(\phi)=\frac{1}{6}\xi^2(t)\phi\left[\tau(t)
+\frac{1}{\lambda(t)}\xi(t)^2\phi^2\right] = \frac{J}{\sqrt{\lambda}} \, .
\end{equation}
We can also determine the second derivative of $ V_{RGI}(t)$ which, after recalling 
the definition of $J$ in Eq. (\ref{dpotimpr}), reads
\begin{equation}
\label{dspotimpr}
 V''_{RGI} = \frac{J}{\phi  \sqrt{\lambda} } +
\frac{1}{3\lambda(t)}\xi(t)^4\phi^2.
\end{equation}

\step

Let us now go back to a somewhat delicate point.
The model under investigation has two characteristic scales
 $M_1(\phi)$  and $M_2(\phi)$,  respectively associated with the 
mass of the longitudinal field and with the  Goldsone bosons mass:
$M_1^2(\phi)=\lambda V''(\phi)$ and 
$M_2^2(\phi)=\lambda V'(\phi)/ \phi$.
If these two scales are comparable, then the usual procedure to compute
$V_{RGI}(\phi)$, that we have outlined above, is essentially correct.
On the other hand, when these scales are significantly decoupled, 
new aspects, that we have overlooked up to now, appear \cite{Ford:1996hd}.
It is important to observe that  some perturbative
diagrams contributing to the longitudinal two point function
 are affected by infrared  divergences due to the loops of Goldstone bosons 
and a reliable estimate of $V''_{RGI}(\phi)$ is obtained only when the  
Goldstone contributions are included in the evaluation of the Renormalization Group  functions.  
Indeed, the Goldstone bosons have zero mass in absence of external source
and, when $J$ is turned on, their square mass is proportional to $J$, while the longitudinal field 
starts with a nonvanishing bare mass at $t=0$ and, at least in the $N\to \infty$ case, 
its renormalized square mass is proportional to  $J^{\epsilon/2}$ when $J$ approaches zero. 
Therefore, in the evaluation of the  $\beta$-function 
we expect that the deep infrared region is dominated by the smaller scale 
$M_2(\phi)$, so that the contribution of the larger scale  $M_1(\phi)$ can be neglected.
Accordingly, we select $\mu^2=M^2_2(\phi)=J\sqrt{\lambda}/\phi$
with $t=\ln(\mu /\Lambda)$ and  Eqs. (\ref{beta},\ref{betam}) are replaced by
\begin{eqnarray}
\label{beta1}
\frac{du}{dt} &=& -\epsilon u+ C_d \frac{N-1}{3}u^2,\\
\label{beta1m}
\frac{d (M^2-M_c^2)}{dt}&=& C_d \frac{N-1}{3} u (M^2-M_c^2) ,
\end{eqnarray}
Consequently Eq. (\ref{betat}) becomes $d \tau/ d t =0$  so that its solution
in (\ref{tausol}) now becomes $\tau(t)=\tau$, while the factor $(N+8)$ has to be replaced by $(N-1)$ 
in Eqs. (\ref{betaphi},\ref{usol},\ref{lambdasol})
and Eq. (\ref{xisol}) is formally unchanged.

\step 

Since our analysis concerns the infrared region corresponding to large negative $t$,
at this point it is essential to recall the different behavior of the theory in $2<d<4$ and $d=4$,
due to the presence, in the former case, of the  Wilson-Fisher fixed point 
which merges with the Gaussian fixed point when $d=4$.
Therefore we have to treat the two cases separately and we start by considering $2<d<4$. 
Then, from Eq. (\ref{beta1}) one finds the Wilson-Fisher  fixed point 
\begin{equation}  
\label{wisfish}   
u^*=\frac{3 \epsilon}{C_d (N-1)}   
\end{equation}   
(that vanishes when $\epsilon\to 0$)
and the scaling of $u(t)$ and $\xi(t)$ near this fixed point is
\begin{equation}
\label{couinfra}
\frac{1}{u(t)} \simeq \frac{1}{u^*}+ \left ( \frac{1}{u} - \frac{1}{u^*} \right ) e^{\epsilon t}
\end{equation}
\begin{equation}
\label{xiinfra}
\xi(t)\simeq\left(\frac{u}{u^*}\right)^{-1/2}e^{\frac{\epsilon}{2}t} \, .
\end{equation}
From the infrared  behavior of the coupling in Eq.  (\ref{couinfra}), 
one can write  for the corresponding dimensionful quantity, after  neglecting subleading terms:
\begin{equation}
\label{laminfra}
\frac{\lambda}{\lambda(t)}\simeq\frac {u}{u^*} 
e^{- \epsilon t} \, .
\end{equation}

\step 

Now we can focus on the 
inverse susceptibility which, due to the rescaling of  $\phi$ in Eq. (\ref{newfi}),
is $\chi_L^{-1}(J)=\sqrt{\lambda}\,( \partial J /\partial \phi ) = \lambda  V''_{RGI}$
and it can be read from Eq. (\ref{dspotimpr}) after inserting the explicit form of the running parameters. 
Then, according to Eqs. (\ref{xiinfra}, \ref{laminfra}), 
for small values of the source the linear term in $J$ in  the right hand side of Eq. 
(\ref{dspotimpr}) can be neglected with respect to the other one, and  one finds:
\begin{equation}
\label{inversesusc}
 \chi_L^{-1}(J)=\lambda  V''_{RGI} \simeq \frac{u^*}{3u}  e^{\epsilon t}  \phi^2=
\frac{u^*}{3u}  \left ( \frac{ J \sqrt{\lambda} }{\phi \Lambda^2} \right)^{(4-d) /2}  \phi^2
\end{equation}
In the limit $J\to 0$ in Eq. (\ref{inversesusc}), the field  $\phi$
can be replaced with $\phi_0$ (up to subleading terms), so the inverse susceptibility vanishes as
$\chi_L^{-1}(J)\propto J^{\epsilon/2}$ when $J\to 0$, which is the same behavior found in 
Eq. (\ref{suscdependence})  in the limit of large $N$.

\step

The comparison with the large $N$ case can be pushed forward by looking at 
Eq. (\ref{dpotimpr}) and by
including the scale dependence displayed in 
Eqs. (\ref{wisfish},\ref{couinfra},\ref{xiinfra},\ref{laminfra}). For large $N$ we get
\begin{equation}
\label{confronto2}
 \frac{J}{\sqrt{\lambda}}=\frac{\phi}{6} \left (\tau +\frac{\phi^2}{\lambda} \right )
\left [ \left(\frac{  J\sqrt{\lambda} }{\phi\Lambda^2}\right)^{(4-d)/2} \frac{u^*}{u} \right ]=
\frac{\phi}{6\lambda} \left (\tau +\frac{\phi^2}{\lambda} \right )
\left[\frac{3 \epsilon}{C_d \, N}\left(  \frac{J \sqrt{\lambda} }{\phi}  \right)^{(4-d)/2}\right].
\end{equation}
It is not difficult to see that Eq.\,(\ref{confronto2}) coincides
with the  $N\to \infty$ result in Eq. (\ref{sadagain}), at least in the 
region of small $J$ and 
$0<\epsilon<<1$, where one can neglect the linear term in $\rho$ and 
approximate $\Gamma(1-d/2)=\Gamma(-1+\epsilon/2)\simeq - 2/\epsilon$.
Incidentally, we notice that the particular choice of the Goldstone mass as the infrared scale and of the $\beta$-function in 
(\ref{beta1}), has selected only the Goldstone bosons contribution,
which is also dominant  in the large $N$ approximation.
In this sense,  it is not surprising to find the same result in Eqs. (\ref{sadagain}) and  (\ref{confronto2}).

\step

Finally we analyze the problem in $d=4$, i.e. $\epsilon = 0$ , where 
the coupling is  dimensionless ($\lambda=u$)  and,
as is well known and as observed in the
$N\to \infty$ case in Sec \ref{largen}, 
the divergences associated  to the $\epsilon^{-1}$ poles 
are typically  replaced by logarithms of the Renormalization Group scale.
In addition, the main difference with respect to the problem in $d<4$
is that the Wilson-Fisher and the Gaussian fixed points coincide in $d=4$, $u^*=0$,
and this in turn implies that, when the ultraviolet cutoff is removed, one ends up 
with  a noninteracting theory. In practice we can observe this 'triviality' property 
directly from the coupling constant, as obtained by integrating Eq. (\ref{beta1}) 
in the limit  $\epsilon \to   0$ (below we define $h= \frac{u}{16\pi^2}\frac{N-1}{3}$)
\begin{equation}
\label{u4d}
u(t)=\frac{u}{1- h \, t}\sim\frac{u}{h\,|t|} \, \, ,
\end{equation}
where the last term is obtained in the region of  large negative $t$.
In fact, from the second term in  Eq. (\ref{u4d})  (i.e. before considering its approximation for large negative $t$ 
that is displayed in the right hand side of Eq. (\ref{u4d})), 
it is evident that the limit of infinite  ultraviolet 
cutoff $\Lambda\to \infty$, with fixed  bare coupling $u$,  
corresponds to  $t\to - \infty$ for any finite infrared scale $\mu$ 
and therefore one finds $u(t=- \infty)=0$, i.e. the coupling vanishes at any finite $\mu$.
At the same time, we notice that the noninteracting limit  $u(t=- \infty)=0$ is recovered also 
at finite $\Lambda$, but  for a vanishing infrared scale $\mu=0$ and, in conclusion, 
a nonvanishing coupling $u(t)\neq 0$ requires finite values of $t$, i.e. finite infrared and ultraviolet scales.

Then, going back to our problem,  from  
Eqs. (\ref{beta1},\ref{beta1m},\ref{xisol}), in $d=4$ and  for large negative but finite  $t$, we find 
$\tau(t)=\tau$ and $\xi(t)\sim\left( h\, |t|\right)^{-1/2}$. 
With these results we can immediately specialize Eq. (\ref{dpotimpr}) to $d=4$:
\begin{equation}
\label{dpotd4}
\frac{J}{\sqrt{u}}=
\frac{\phi}{6}\left [\tau \left( h\, |t| \right)^{- 1}
+ \frac{\phi^2}{u} \, \left( h\, |t| \right)^{-1} \right ] \,\, ,
\end{equation}
and, in the same way,  the second derivative of the potential in Eq. (\ref{dspotimpr}) becomes
\begin{equation}
\label{dspotd4}
\chi_L^{-1} (J) = u \, V''_{RGI}=\frac{\sqrt{u} J}{\phi} + \frac{\phi^2}{3 h |t|} \,\, ,
\end{equation}
where  $J$ has been inserted  according to Eq. (\ref{dpotd4}).
Eq. (\ref{dspotd4}) shows that the inverse longitudinal susceptibility, as obtained from $V_{RGI}$, 
depends only on the ratio of  the Renormalization Group scales $\Lambda$ and $\mu$ through the logarithm $t$.
Then,  since the infrared scale is taken equal to the Goldstone boson mass, $\mu^2=M^2_2(\phi)=J\sqrt{\lambda}/\phi$,
we recover the logarithmic dependence of $\chi_L^{-1} (J)$ on the source $J$
(after neglecting the first term in the right hand side of  Eq. (\ref{dspotd4})  which vanishes linearly
with $J$).  It is also straightforward to realize that in the large $N$ limit,  Eq. (\ref{dspotd4})
reproduces Eq.(\ref{susc}) with the loop integral displayed in  Eq. (\ref{iduequattro}).

\step

Therefore, our conclusion in $d=4$ is that $\chi_L^{-1} (J)$ vanishes as $(\ln ( J )  )^{-1}$
when the external source is turned off, exactly as it happens in the large $N$ limit.
However it must be remarked that, because of the  logarithmic dependence of our observable on the 
ratio of two scales, the limit $J\to 0$ cannot be 
disentangled from the limit $\Lambda \to \infty$ and this means that the 
results obtained in the limit  $J\to 0$ are those of  a noninteracting theory. 

\step

Before finishing we would like to investigate the possibility of observing a nonvanishing longitudinal susceptibility  in $d=4$.  
According to the above analysis, it is clear that formally the only way of achieving this result is to stay with a finite value of 
$t$, by keeping a finite $\Lambda$ and also a finite infrared scale 
which for the moment could be taken as a  generic  value of the field:   $\mu^2=\phi^2/u$. If $\mu$ is taken much 
larger than  $M_1$ and $M_2$ introduced above,  then the problem  related to the presence of two different 
infrared scales is no longer present and we can use  the $\beta$-functions in Eqs. (\ref{beta}) and (\ref{betam}). 
This leads to a rescaling of the factor 
$h\to h_r=h(N+8)/(N-1)$ in Eq. (\ref{u4d}) and to the  following explicit form of Eqs. (\ref{tausol}) and (\ref{xisol}):
\begin{equation}
 \tau(t)\sim\tau\left(h_r\, |t|\right)^{\frac{6}{N+8}}
\;\;\;\;\;\;\;  ; \;\;\;\; \;\;\;
 \xi(t)\sim\left( h_r\, |t|\right)^{-1/2} \, .
\end{equation}
As a consequence Eq. (\ref{dpotimpr}) now becomes 
\begin{equation}
\label{dpotd4bis}
\frac{J}{\sqrt{u}}=
\frac{\phi}{6}\left [\tau \left( h_r\, |t| \right)^{- \frac{N+2}{N+8}}
+ \frac{\phi^2}{u} \, \left( h_r\, |t| \right)^{-1} \right ] \, .
\end{equation}
The minimum of $V_{RGI}$ is located at $\phi_0\neq 0$, such that $J (\phi_0)= 0$ in Eq. (\ref{dpotd4bis}) :
\begin{equation}
\label{phinought}
\phi_0^2=-\tau \, u  \left( \,\frac{h_r}{2} \,\, \Big| \ln\left (\frac{\phi^2_0}{u\Lambda^2}\right )
\Big| \, \right)^{\frac{6}{N+8}} \, .
\end{equation}
\step 
The square magnetization $\phi^2_0/u$ displayed in Eq. (\ref{phinought}), or equivalently the temperature $\tau$,
corresponds to a finite quantity that  can be used  to set the infrared scale in this calculation.  
Then Eq. (\ref{dspotimpr}), computed at  $\phi=\phi_0$ (i.e. at $J=0$) and with $\phi_0$ replaced by $\tau$ according to  
Eq. (\ref{phinought}), gives (up to  logarithmic corrections), \cite{zinn}, 
\begin{equation}
\label{chiinvefin}
\chi_L^{-1}\simeq -\frac{\tau u}{3} \left( \,\frac{h}{2} \,\, \Big| \ln\left (\frac{-\tau}{\Lambda^2}\right )
\Big| \, \right)^{-\frac{N+2}{N+8}} \, .
\end{equation}
Eq. (\ref{chiinvefin}) provides a nonvanishing  $\chi_L^{-1}$, but  it is evident that this  result  can only  be obtained at 
the price of constraining $t$  to a finite nonvanishing value,  which in turn means that we are dealing with an interacting 
theory, effectively defined for values of the  momentum smaller than $\Lambda$,  and where  the effects of the Goldstone 
bosons are  screened by  the finite  infrared scale.  

\step 

\section{Conclusions}
\label{discussion}

We studied the longitudinal susceptibility in the broken phase of the $O(N)$ theory, both in the simpler case of the large $N$ 
limit and in the general case of finite $N$. In this analysis we compared the results of three different approaches, 
namely the leading contribution of the $1/N$ expansion, the Functional Renormalization Group flow equations in the Local 
Potential approximation and the improved effective potential via the Callan-Symanzik equations, properly extended to 
four dimensions through the  expansion in powers of $\epsilon$.

\step

In the large $N$ limit, the Functional Renormalization Group correctly reproduces a convex potential with a flat part in the
region $|\Phi|\leq |\Phi_0|$, while this does not occur in the potential obtained in the large $N$ expansion. In fact, 
the latter potential is computed by performing an asymptotic expansion of the path integral in $Z[J]$ around a single 
saddle point configuration.
In order to recover the convexity of the effective potential  for   $|\Phi|\leq |\Phi_0|$,
the contribution of the other saddle points has to be considered.  
The virtue of the Functional Renormalization Group is that it takes into account by  construction all of these contributions. 
On the other hand, for  $|\Phi|\geq |\Phi_0|$, the agreement  of the two approaches is excellent.  

\step

According to the saddle point analysis performed in the large $N$ limit , the inverse longitudinal  susceptibility 
$\chi_L^{-1}$ vanishes  with the external source $J$ as $\chi_L^{-1}\propto J^{\epsilon/2}$, for $2<d<4$.
We have also shown that this scaling law is limited to a very narrow  region around $J=0$. 
In $d=4$ the power law  is replaced  by the logarithm of the ratio of the ultraviolet cutoff and the infrared scale 
associated to the source $J$ and  $\chi_L^{-1}(J)$  vanishes with $J$ as $(\ln (J))^{-1}$.

\step 

At finite $N$,  the Functional Renormalization Group numerical analysis  shows  the vanishing of $\chi_L^{-1}$  for $d=3$ and $d=3.5$.
In $d=4$ the region around the minimum of the potential is affected by strong numerical fluctuations
and, as a consequence, it is not possible to conclude from the numerical analysis whether or not  $\chi_L^{-1}$ has a discontinuity at $|\phi_0|$.
This point is clarified by means of  the Callan-Symanzik approach. Within this latter framework we analyzed  the longitudinal susceptibility for 
$\epsilon\geq 0$ and recovered the same behavior already found in the limit  $N\to\infty$, namely $\chi^{-1}_L$ vanishing with $J$
as $J^{\epsilon/2}$ for $\epsilon>0$ and as $(\ln (J))^{-1}$ in $d=4$, in agreement with the results of \cite{Anishetty:1995kj}. 

\step 

More generally, we find a full correspondence of our Renormalization Group  improved expressions of the derivatives of the effective potential
in the Callan-Symanzik approach when $N\rightarrow \infty$, with the same quantities computed at the leading order in the  large $N$ expansion.
Finally, in this approach it is clearly illustrated how the behavior of the inverse longitudinal susceptibility in $d=4$ is simultaneously  related to 
the vanishing of the Goldstone boson mass and to the limit $\Lambda \to \infty$ which leads to a noninteracting scalar theory.

\end{document}